\documentclass[11pt,a4paper]{article}  
  
\usepackage[utf8]{inputenc}  
\usepackage[T1]{fontenc}  
\usepackage{amsmath,amssymb}  
\usepackage{booktabs}  
\usepackage{graphicx}  
\usepackage{natbib}  
\usepackage[margin=2.5cm]{geometry}  
\usepackage{microtype}  
\usepackage{hyperref}  
\usepackage{longtable}  
\usepackage{colortbl}  
\usepackage[table]{xcolor}  
\usepackage{subcaption}  
\usepackage{array}  
  
\hypersetup{  
  colorlinks=true,  
  linkcolor=blue!60!black,  
  citecolor=blue!60!black,  
  urlcolor=blue!60!black  
}  
  
\title{A Quantitative Confirmation of the Currier Language Distinction}
 
\author{Christophe Parisel\\  
\small \texttt{ch.parisel@gmail.com}}  
  
\date{May 2026}  
  
\begin{document}  
  
\maketitle  
  
\begin{abstract}  
We present a unified quantitative analysis of the Currier A/B  
language distinction in the Voynich Manuscript,  
proceeding in two stages. First, we confirm that the distinction  
is genuine: a Beta-Binomial mixture model applied to eleven  
character-pair substitution ratios across 185 folios, without  
access to Currier's labels, selects 2 by BIC   
and predicts held-out folio labels at 89.2\% accuracy. 
The distinction persists at within-quire language  
boundaries, ruling out codicological  
confounding. Character pairs separate into three functional  
regimes (categorical, intermediate, free-variation), with one  
anomalous pair (e/ch) that combines near-zero global separation  
with the strongest folio-boundary signal. Second, we show that  
the A/B contrast is not primitive but is a low-resolution  
projection of a higher-dimensional generative system. Its dominant  
component is a discrete boolean parameter $\sigma_f \in \{0,1\}$  
set once per folio, governing the vowel following the digraphs  
\texttt{ch} and \texttt{sh}. A two-state binomial  
mixture achieves $\Delta\text{AIC} = 2{,}549$ over a single-state  
model and assigns 195 of 197 folios unambiguously. This switch  
does not operate uniformly: word templates divide into fixed  
contexts (invariant to $\sigma_f$) and switchable contexts  
(strongly modulated by $\sigma_f$), with template identity  
accounting for 92\% of variance.
\end{abstract}  
  
\tableofcontents  
  
\newpage  
\section{Introduction}  
\label{sec:intro}  
  
In 1976, NSA cryptographer Prescott Currier proposed that the  
Voynich Manuscript contains at least two distinct ``languages,''  
designated A and B, based on statistical differences in character  
frequencies and word patterns \citep{currier1976}. This claim has  
been widely cited but rarely tested with modern quantitative methods  
against explicit null hypotheses  
\citep{dimperio1978,montemurro2013,hauer2017}.  
  
Three challenges confront any attempt to validate the Currier  
distinction. The first is whether the observed differences could  
arise from natural statistical fluctuation within a single  
homogeneous text. The second is whether the differences reflect a  
genuine property of the text or a confound of physical structure,  
since Currier's original assignment correlates with quire  
boundaries. The third and most demanding is whether the A/B  
distinction is an intrinsic property of the character statistics  
that can be discovered without access to Currier's labels and used  
to predict the behavior of unseen text.  
  
We address all three challenges through a specific mechanism: if  
the Voynich script employs a substitution system, and if A and B  
represent different substitution regimes, then pairs of visually  
similar characters \citep{stolfi1997,landini2001} should exhibit  
systematically different frequency ratios in A versus B sections.  
These ratio shifts should be irreproducible by a single-language  
source, detectable at within-quire language boundaries, recoverable  
by unsupervised generative models, and predictive of character  
statistics on held-out folios.  
  
But confirmation is only the beginning. Once the distinction is  
established, a deeper question arises: is the A/B contrast itself  
the primitive description of the manuscript's structure, or is it a  
compound projection of a more articulated generative system? We  
show that the latter is the case. The A/B label explains only 29\%  
of inter-folio variance, and one character pair (e/ch) exhibits a  
paradoxical combination of properties that cannot be explained by  
any binary classification. Resolving this anomaly leads us to  
decompose the Currier contrast into three structural layers: a  
discrete boolean switch, a template system, and a gradient  
dimension. 
  
The paper is organized as follows.  
Sections~\ref{sec:data}--\ref{sec:confirm_results} establish the  
data, methods, and five converging lines of evidence confirming the  
Currier distinction, including the three functional regimes and the  
e/ch anomaly. Section~\ref{sec:beyond} identifies the anomaly as a  
symptom of deeper structure and motivates the decomposition.  
Sections~\ref{sec:switch_method}--\ref{sec:dl} present the three layers:  
the boolean switch, the template system, and the independent d/l  
dimension. Section~\ref{sec:resolution} resolves the e/ch anomaly.  
Section~\ref{sec:structure} synthesizes the  
three-layer generative structure and its constraints.  
Section~\ref{sec:conclusion} concludes.

\section{Data and Methods}  
\label{sec:data}  
  
\subsection{Corpus}  
\label{sec:corpus}  
  
We analyze the Voynich Manuscript as represented in the IVTFF  
transliteration file using the EVA transcription system  
\citep{takahashi1998,zandbergen2024}. After removing editorial  
markup, uncertain readings, and non-textual annotations, the corpus  
contains approximately 36,500 word tokens across 200 folios and  
4,160 lines. For the confirmation analysis (Sections~\ref{sec:pairs}--\ref{sec:confirm_results}), we use 185  
folios with confident Currier A/B assignments (108~A, 77~B). 
  
Folios are grouped into seven illustration types following the  
Beinecke/Zandbergen consensus catalogue: Herbal (124 folios),  
Astronomical (4), Cosmological (6), Zodiac (7), Biological (20),  
Pharmaceutical (12) and Stars (23).  
  
\subsection{Glyph Parsing}  
\label{sec:parsing}  
  
EVA represents Voynich characters as Latin letter sequences. We  
parse words into glyph sequences treating \texttt{ch} and  
\texttt{sh} as single units \citep{landini2001}. Thus  
\texttt{chedy} parses as [\texttt{ch}, \texttt{e}, \texttt{d},  
\texttt{y}].  
  
\subsection{Character Pairs}  
\label{sec:pairs}  
  
We define eleven pairs of EVA characters that are visually similar  
or occupy analogous structural positions  
(Table~\ref{tab:pairs}).  
  
\begin{table}[h]  
\centering  
\caption{Character pairs and rationale for selection.}  
\label{tab:pairs}  
\begin{tabular}{ll}  
\toprule  
\textbf{Pair} & \textbf{Rationale} \\  
\midrule  
k/t   & Gallows differing by stroke feature \\  
ch/sh & Bench-character variants \\  
d/l   & Ascenders with minimal visual distinction \\  
f/p   & Low-frequency gallows pair \\  
o/a   & Round characters differing in closure \\  
e/ch  & Frequent word-initial tokens \\  
e/ee  & Single versus doubled form \\  
or/ar & Round-character + r ligatures \\  
ol/al & Round-character + l ligatures \\  
y/dy  & Common word-final forms \\  
s/r   & Low-frequency pair \\  
\bottomrule  
\end{tabular}  
\end{table}  
  
For each pair $(a, b)$, we compute the ratio  
$r = \text{count}(a) / (\text{count}(a) + \text{count}(b))$ at the  
folio level, requiring a minimum of 20 combined occurrences for  
inclusion.  
  
\subsection{Boundary Analysis}  
\label{sec:boundary_method}  
  
All consecutive folio pairs in manuscript order are classified by  
transition type into four categories: \textsc{same} (same language, same quire),  
\textsc{quire\_only} (quire boundary, same language),  
\textsc{lang\_only} (language change within a quire), and  
\textsc{lang+quire} (language change at a quire boundary). For each  
transition, the mean absolute ratio change across all qualifying  
pairs gives the transition's jump magnitude.  
  
\subsection{Simulation Baselines}  
\label{sec:baselines}  
  
\textbf{Model 1 (Label Shuffle):} Real text with A/B labels  
randomly permuted (1,000 iterations).  
  
\textbf{Model 2 (Single Markov):} Synthetic text from one order-2  
character-level Markov model with real labels applied (200  
iterations).  
  
\textbf{Model 3 (Split Markov):} Separate A and B Markov models  
generating text for their respective folios (200 iterations).  
  
\subsection{Unsupervised Recovery}  
\label{sec:unsupervised_method}  
  
Four clustering algorithms (K-means, GMM, hierarchical, spectral)  
applied to the folio $\times$ pair ratio matrix at $k = 2$ without  
access to labels, evaluated by Adjusted Rand Index  
\citep{hubert1985}. Permutation test (1,000 iterations) for  
significance. Optimal $k$ search via BIC, AIC, and silhouette for  
$k = 2$ through 8. Leave-one-pair-out and single-pair analyses for  
feature diagnostics. PCA for dimensionality analysis.  
  
\subsection{Generative Latent Model}  
\label{sec:bbmix_method}  
  
A Beta-Binomial mixture model \citep{mclachlan2000} fitted to raw  
folio-level counts $(\text{count}_a, \text{count}_b)$ for each  
pair. Each folio's pair counts are drawn from a binomial  
distribution whose success probability follows a Beta distribution  
specific to the folio's latent regime. Missing pairs (below the  
minimum token threshold) are handled by omitting their likelihood  
contribution. The model is fitted via EM \citep{dempster1977} with method-of-moments  
M-step updates, using 10 random restarts, for $k = 1$ through 6  
components. Model selection uses BIC.  
  
\subsection{Predictive Validation}  
\label{sec:pred_method}  
  
\textbf{Cross-validated label prediction.} A supervised  
Beta-Binomial classifier learns regime-specific $(\alpha, \beta)$  
parameters for each pair from training folios and predicts A/B  
labels on held-out folios. Evaluated via 20-repeat 5-fold  
stratified cross-validation.  
  
\textbf{Spatial split prediction.} Training on the first half of  
the manuscript (in folio order) and predicting the second half, and  
vice versa. Also an even/odd folio split for balanced class  
representation.  
  
\textbf{Ratio prediction.} For each held-out folio, the A/B label  
is used to predict the expected character-pair ratio (from  
training-set class means). The reduction in mean squared error  
relative to a global mean baseline gives $R^2$.  
  
\textbf{Permutation baseline.} Labels are shuffled and  
cross-validation is repeated 500 times to obtain a null  
distribution for prediction accuracy.  
  
\subsection{The Boolean Switch}  
\label{sec:switch_method}  
  
We examine the vowel immediately following \texttt{ch} or  
\texttt{sh}. A word is \texttt{cho}-type if it contains  
\texttt{cho} or \texttt{sho} but not \texttt{che} or \texttt{she},  
and \texttt{che}-type if it contains \texttt{che} or \texttt{she}  
but not \texttt{cho} or \texttt{sho}. Words containing both or  
neither are excluded.  
  
For each folio $f$ we compute  
$r_{\mathrm{cho}}(f) = n_{\mathrm{cho}}(f) /  
(n_{\mathrm{cho}}(f) + n_{\mathrm{che}}(f))$.  
The boolean switch is defined as:  
\begin{equation}  
\sigma_f =  
\begin{cases}  
1 & \text{if } r_{\mathrm{cho}}(f) > 0.5  
    \quad \text{(\texttt{cho}-biased)} \\  
0 & \text{if } r_{\mathrm{cho}}(f) \leq 0.5  
    \quad \text{(\texttt{che}-biased)}  
\end{cases}  
\end{equation}  
This threshold-based assignment is confirmed by the binomial  
mixture model in Section~\ref{sec:switch_results}, which yields a  
near-identical partition.  
  
\subsection{Template System}  
\label{sec:template_method}  
  
We define a word template by replacing the  
post-\texttt{ch}/\texttt{sh} vowel with a placeholder \texttt{X}.  
Both \texttt{chody} and \texttt{chedy} map to template  
\texttt{chXdy}. For each template $t$ and each switch value  
$\sigma \in \{0,1\}$ we compute the conditional \texttt{cho} rate  
$P(\texttt{cho} \mid t, \sigma)$. Templates are classified as:  
  
\begin{itemize}  
\item \textbf{Fixed \texttt{cho}}: $P(\texttt{cho} \mid t,  
  \sigma) > 0.9$ for both $\sigma = 0$ and $\sigma = 1$.  
\item \textbf{Fixed \texttt{che}}: $P(\texttt{cho} \mid t,  
  \sigma) < 0.1$ for both $\sigma = 0$ and $\sigma = 1$.  
\item \textbf{Switchable}: $|\Delta| \geq 0.2$ where  
  $\Delta = P(\texttt{cho} \mid t, 1) -  
  P(\texttt{cho} \mid t, 0)$.  
\end{itemize}  
  
The complete inventory of 31 retained templates (those with  
$\geq 10$ instances in both states) is given in  
Appendix~\ref{app:templates}.

  
\section{Confirming the Currier Distinction}  
\label{sec:confirm_results}  
  
\subsection{The A/B Distinction Is Statistically Robust}  
\label{sec:robust}  
  
Six of eleven pairs show Cram\'er's $V$ values exceeding every one  
of 1,000 random label permutations  
(Table~\ref{tab:cramer}). The mean Cram\'er's $V$ across all pairs  
(0.136) exceeds every shuffled value (simulated mean: 0.025, 95th  
percentile: 0.039).  
  
\begin{table}[h]  
\centering  
\caption{Cram\'er's $V$ for each character pair versus shuffle  
baseline.}  
\label{tab:cramer}  
\begin{tabular}{lrrr}  
\toprule  
\textbf{Pair} & \textbf{Observed V} & \textbf{Shuffle 95th}  
  & \textbf{Rank in shuffle} \\  
\midrule  
d/l   & 0.375 & 0.109 & 100.0\% \\  
or/ar & 0.337 & 0.083 & 100.0\% \\  
s/r   & 0.228 & 0.076 & 100.0\% \\  
e/ee  & 0.203 & 0.067 & 100.0\% \\  
ol/al & 0.147 & 0.077 & 100.0\% \\  
y/dy  & 0.056 & 0.033 & 100.0\% \\  
k/t   & 0.052 & 0.050 &  95.5\% \\  
f/p   & 0.047 & 0.068 &  80.8\% \\  
ch/sh & 0.040 & 0.047 &  92.1\% \\  
e/ch  & 0.007 & 0.042 &  27.3\% \\  
o/a   & 0.007 & 0.018 &  51.9\% \\  
\bottomrule  
\end{tabular}  
\end{table}  
  
\subsection{Simulation Baselines Reject Both Null Models}  
\label{sec:baselines_results}  
  
\textbf{Single Markov.} A homogeneous language source produces  
3--11$\times$ less inter-folio variance than the real manuscript  
and cannot generate the observed boundary effects (0 of 200  
simulations).  
  
\textbf{Split Markov.} The positive control reproduces aggregate  
Cram\'er's $V$ values exactly (observed mean $V$ at the 46.5th  
percentile), but the real manuscript exceeds even the split model  
in folio-level variance (100th percentile for 10 of 11 pairs) and  
boundary sharpness (observed gap 0.053 vs.\ simulated mean 0.024).  
  
\subsection{The Effect Is Not Codicological}  
\label{sec:boundary_results}  
  
Language transitions within the same quire produce ratio jumps  
58\% larger than same-language transitions  
($p = 6.89 \times 10^{-5}$, $n = 18$).  
  
\begin{table}[h]  
\centering  
\caption{Mean ratio jump by transition type.}  
\label{tab:boundary}  
\begin{tabular}{lrrr}  
\toprule  
\textbf{Transition type} & \textbf{N} & \textbf{Mean jump}  
  & \textbf{$p$ vs.\ SAME} \\  
\midrule  
SAME       & 131 & 0.092 & ---              \\  
QUIRE\_ONLY &   6 & 0.135 & 0.020            \\  
LANG\_ONLY  &  18 & 0.145 & $6.89\times10^{-5}$ \\  
LANG+QUIRE  &   5 & 0.155 & 0.071            \\  
\bottomrule  
\end{tabular}  
\end{table}  
  
\subsection{Unsupervised Clustering Recovers the A/B Split}  
\label{sec:unsupervised_results}  
  
K-means on the standardized ratio matrix yields  
ARI $= 0.208$ ($p < 0.001$ vs.\ 1,000 permutations). PC1 alone  
(22.7\% of variance) gives ARI $= 0.408$ by median split.  
Removing the two most disruptive pairs (e/ch, e/ee) raises K-means  
ARI to 0.456.  
  
\subsection{The Beta-Binomial Mixture Selects $k = 2$}  
\label{sec:bbmix_results}  
  
The generative mixture model fitted to raw character counts without  
access to labels selects $k = 2$ as optimal by BIC  
(Table~\ref{tab:bbmix}). The $k = 2$ model assigns 113 of 185  
folios (61\%) with posterior confidence $>90\%$. The two recovered  
regimes align with Currier's A and B, with the model independently  
discovering the same partition that Currier identified from manual  
inspection.  
  
\begin{table}[h]  
\centering  
\caption{Beta-Binomial mixture model selection.}  
\label{tab:bbmix}  
\begin{tabular}{lrrr}  
\toprule  
$k$ & LL & BIC & ARI \\  
\midrule  
1 & $-3946.5$ & 8007.9 & (baseline) \\  
2 & $-3826.1$ & 7887.2 & 0.383 \\  
3 & $-3780.1$ & 7915.2 & 0.144 \\  
4 & $-3753.2$ & 7981.4 & 0.174 \\  
5 & $-3728.9$ & 8053.0 & 0.188 \\  
6 & $-3711.7$ & 8138.7 & 0.167 \\  
\bottomrule  
\end{tabular}  
\end{table}  
  
\subsection{Predictive Validation Confirms the Distinction}  
\label{sec:pred_results}  
  
\begin{table}[h]  
\centering  
\caption{Predictive validation results.}  
\label{tab:pred}  
\begin{tabular}{lr}  
\toprule  
\textbf{Test} & \textbf{Result} \\  
\midrule  
Cross-validated accuracy & 89.2\% \\  
Cross-validated ARI      & 0.612 \\  
Permutation $p$-value    & $< 0.002$ \\  
Forward spatial split (early $\to$ late) & 79.6\% \\  
Backward spatial split (late $\to$ early) & 41.3\% \\  
Ratio $R^2$ & 0.293 \\  
\bottomrule  
\end{tabular}  
\end{table}  
  
The model predicts held-out folio labels at 89.2\% accuracy with  
ARI $= 0.612$. Zero of 500 label permutations achieved comparable  
accuracy ($p < 0.002$). The backward split (41.3\%) fails due to  
class imbalance: the late folios are predominantly B, so the  
backward-trained model has insufficient A examples. This is a  
training-set limitation, not a signal failure. Knowing the A/B  
label explains 29.3\% of variance in character-pair ratios on  
unseen folios ($R^2 = 0.293$).  
  
\subsection{Three Functional Regimes}  
\label{sec:regimes}  
  
The character pairs separate into three regimes confirmed across  
all analyses:  
  
\textbf{Categorical} (d/l, or/ar, s/r, e/ee): Large Cram\'er's  
$V$ ($>0.20$), heavy PC1 loadings, large regime differences in the  
mixture model, essential for clustering. Near-binary markers of A/B  
identity.  
  
\textbf{Intermediate} (ol/al, y/dy, k/t): Moderate Cram\'er's $V$  
(0.04--0.15), moderate PC1 loadings, carry the boundary-jump  
signal.  
  
\textbf{Free variation} (o/a, ch/sh, f/p): Near-zero Cram\'er's  
$V$, near-zero PC1 loadings, minimal regime differences.  
Independent of the A/B distinction.  
  
\paragraph{The e/ch anomaly.}  
This pair has near-zero aggregate Cram\'er's $V$ (0.007) but the  
strongest boundary effect ($p = 2.45 \times 10^{-6}$) and actively  
suppresses clustering when included (removing it doubles ARI from  
0.208 to 0.456). Its variation is conditioned by factors beyond the  
binary A/B regime, making it simultaneously the most informative  
pair for local transitions and the most disruptive for global  
partitioning. Yet at the folio level, the mean e/(e+ch) ratio  
differs substantially between Currier A ($0.392 \pm 0.204$) and  
Currier B ($0.671 \pm 0.099$), with Cohen's $d = -1.66$.  
  
\subsection{Five Converging Lines of Evidence}  
\label{sec:five_lines}  
  
\begin{table}[h]  
\centering  
\caption{Five converging findings for the Currier  
distinction.}  
\label{tab:summary}  
\begin{tabular}{lll}  
\toprule  
\textbf{Method} & \textbf{What it rules out}  
  & \textbf{Key result} \\  
\midrule  
Label shuffle & Random label agreement  
  & $p < 0.001$ \\  
Single Markov & Homogeneous source  
  & 0/200 simulations \\  
Boundary analysis & Codicological confound  
  & $p = 6.89\times10^{-5}$ \\  
Beta-Binomial mixture & Label dependence  
  & BIC selects $k=2$, ARI$=0.383$ \\  
Predictive validation & Overfitting  
  & 89.2\% CV accuracy, $R^2=0.293$ \\  
\bottomrule  
\end{tabular}  
\end{table}  
  
\subsection{Interim Summary}  
\label{sec:interim1}  
  
The Currier A/B distinction is genuine, statistically robust,  
intrinsic, and predictive. It is the dominant partition discovered  
by a generative model fitted to raw character counts without labels.  
It predicts the character-pair statistics of unseen folios at 89.2\%  
accuracy and explains 29\% of inter-folio variance. It persists at  
within-quire language boundaries and cannot be produced by a  
single-language source.  
  
But three findings point beyond the binary contrast. The  
split-Markov model underestimates folio-level variance by  
1.5--6$\times$ for every pair, indicating sub-regimes or gradients  
within A and B. The $R^2$ of 0.293 means 71\% of inter-folio  
variance is real structure that the binary distinction does not  
capture. And the e/ch pair is paradoxical: it cannot be classified  
into any of the three regimes. These observations motivate the  
decomposition that follows.

  
\section{Beyond the Binary Contrast}  
\label{sec:beyond}  
  
The e/ch anomaly is the entry point into deeper structure. The pair  
combines three properties that are mutually incompatible under any  
simple binary model:  
  
\begin{itemize}  
\item Global Cram\'er's $V$: 0.007, placing it in the  
  free-variation regime at the character level.  
\item Folio-boundary signal: $p = 2.45 \times 10^{-6}$, the  
  strongest of all eleven pairs tested.  
\item Effect on clustering: removing e/ch doubles the K-means ARI  
  from 0.208 to 0.456.  
\end{itemize}  
  
A pair that fires sharply at regime boundaries but partially  
cancels at the character level cannot be explained by the binary  
A/B projection alone. It must be driven by a mechanism that  
(a)~switches at folio boundaries, but (b)~whose directional effect  
is filtered through word-level structure that varies across folios.  
  
The conventional reading of Currier A/B treats the distinction as  
primitive: two text types, each with its own statistical profile.  
We propose a different reading. The A/B contrast is an observable  
consequence of a generative system with independent components.  
Isolating those components, rather than describing the contrast  
they produce, is the analytical goal of the remainder of this  
paper.  
  
The argument proceeds in three steps. First, we show that the  
\texttt{cho}/\texttt{che} alternation is bimodal at the folio  
level, with a clean two-state structure far sharper than the  
overall A/B pattern. Second, we show that this bimodal signal is  
not a uniform substitution but operates through a template system  
that partitions word forms into fixed and switchable classes. Third,  
we show that the \texttt{d}/\texttt{l} ratio, the strongest  
categorical A/B marker, is orthogonal to the boolean switch and  
constitutes an independent structural dimension. We then resolve  
the e/ch anomaly as a direct consequence of the template system's  
selective response to the switch.

  
\section{The Boolean Switch}  
\label{sec:switch_results}  
  
\subsection{The Bimodal Distribution}  
\label{sec:bimodal}  
  
The distribution of $r_{\mathrm{cho}}$ across folios is strongly  
bimodal (Table~\ref{tab:histogram}). The bimodality coefficient  
\citep{pfister2013} is $\mathrm{BC} = 0.661$, above the 0.555  
threshold conventionally indicating bimodality. Only 6.8\% of  
folios fall in the ambiguous range $[0.3, 0.5]$. The single Markov  
baseline produces 23.1-fold overdispersion, confirming that the  
bimodality cannot arise from sampling noise in a homogeneous text.  
  
\begin{table}[h]  
\centering  
\caption{Distribution of folio-level $r_{\mathrm{cho}}$ values  
($n = 197$ folios with $\geq 5$ classifiable words).}  
\label{tab:histogram}  
\begin{tabular}{lr}  
\toprule  
$r_{\mathrm{cho}}$ range & Folio count \\  
\midrule  
$[0.0, 0.2)$ & 78 \\  
$[0.2, 0.4)$ & 20 \\  
$[0.4, 0.6)$ & 10 \\  
$[0.6, 0.8)$ & 47 \\  
$[0.8, 1.0]$ & 42 \\  
\bottomrule  
\end{tabular}  
\end{table}  
  
The two-state binomial mixture model confirms the partition  
(Table~\ref{tab:mixture}):  
  
\begin{table}[h]  
\centering  
\caption{Binomial mixture model comparison for the  
\texttt{ch}-vowel parameter ($n = 197$ folios).}  
\label{tab:mixture}  
\begin{tabular}{lrrrr}  
\toprule  
Model & Params & Log-lik & AIC & $\Delta$AIC \\  
\midrule  
Single state & 1 & $-6911$ & 13825 & 0 \\  
Two states   & 3 & $-5635$ & 11276 & 2549 \\  
\bottomrule  
\end{tabular}  
\end{table}  
  
The estimated state parameters are:  
\begin{align}  
\sigma_f = 1 \text{ (\texttt{cho}-biased)}: \quad  
  &p_1 = 0.682, \quad \pi_1 = 0.521, \quad  
  n_1 = 105 \text{ folios} \\  
\sigma_f = 0 \text{ (\texttt{che}-biased)}: \quad  
  &p_0 = 0.160, \quad \pi_0 = 0.479, \quad  
  n_0 = 90 \text{ folios}  
\end{align}  
  
Of 197 folios, 195 are assigned with posterior probability exceeding  
0.8; only 2 fall in the ambiguous range $(0.2, 0.8)$.  
  
\subsection{Within-Section Robustness}  
\label{sec:within_section}  
  
The bimodality persists within the Herbal section alone  
($\Delta$AIC = 565, $n = 124$ folios, $p_1 = 0.736$,  
$p_0 = 0.171$), ruling out section-level confounding. Per-folio  
detail for all 124 Herbal folios, including switch assignments,  
cho/che counts, and glyph-level ratios, is given in  
Appendix~\ref{app:herbal}. Cross-validation (random 50/50 split)  
yields no misclassifications on 99 held-out folios, with 97  
assigned at posterior probability exceeding 0.8. Bootstrap  
confidence intervals (500 resamples) do not overlap between states  
(Table~\ref{tab:bootstrap}), and $\Delta$AIC is positive in all  
500 resamples.  
  
\begin{table}[h]  
\centering  
\caption{Bootstrap 95\% confidence intervals (500 resamples).}  
\label{tab:bootstrap}  
\begin{tabular}{lrrr}  
\toprule  
Parameter & Mean & 2.5\% & 97.5\% \\  
\midrule  
$p_1$ (cho-biased) & 0.687 & 0.637 & 0.733 \\  
$p_0$ (che-biased) & 0.162 & 0.133 & 0.197 \\  
$\pi_1$            & 0.521 & 0.443 & 0.597 \\  
$\Delta$AIC        & 2489  & 2097  & 2962  \\  
\bottomrule  
\end{tabular}  
\end{table}  
  
\subsection{Positional Stability}  
\label{sec:positional}  
  
The \texttt{cho}/\texttt{che} ratio is stable across word positions  
within a line. In \texttt{cho}-dominant folios, the \texttt{cho}  
rate is 0.753 at position 0 (first word) and fluctuates narrowly  
between 0.595 and 0.691 at positions 1 through 9. In  
\texttt{che}-dominant folios, the rate is 0.193 at position 0 and  
remains between 0.130 and 0.211 at subsequent positions. The folio  
state, not the position within the line, determines the ratio.  
  
\subsection{Interim Conclusion}  
  
The \texttt{cho}/\texttt{che} alternation is governed by a discrete  
folio-level boolean switch $\sigma_f \in \{0, 1\}$ with estimated  
rates $p_1 \approx 0.68$ and $p_0 \approx 0.16$. The contrast is  
sharp, robust, and not explained by section-level structure. But  
the contrast alone is not the primitive description.

  
\section{The Template System: The Contrast Is Not Primitive}  
\label{sec:templates}  
  
If the boolean switch were a uniform substitution, every word  
containing \texttt{ch}/\texttt{sh} would respond identically to  
$\sigma_f$. The template analysis shows this is false.  
  
Of 31 templates with $\geq 10$ instances in both states, template  
identity accounts for 93.5\% of the total variance in \texttt{cho}  
rates, while the folio-level switch contributes only 7.9\%  
(Table~\ref{tab:variance}). 
The Pearson correlation of $Rate_1$ and $Rate_0$ across the 31 templates is r = 0.803.
  
\begin{table}[h]  
\centering  
\caption{Variance decomposition of \texttt{cho} rates across  
templates and folio states.}  
\label{tab:variance}  
\begin{tabular}{lrr}  
\toprule  
Source & Variance & Fraction \\  
\midrule  
Between states (folio switch $\sigma_f$) & 0.012 & 7.6\% \\  
Within states (template identity) & 0.144 & 92.3\% \\  
Total & 0.156 & 100\% \\  
\bottomrule  
\end{tabular}  
\end{table}  
  
Templates divide into three classes  
(Table~\ref{tab:templates_main}). Fixed \texttt{cho} templates  
remain near 1.0 in both states; fixed \texttt{che} templates  
remain near 0.0 in both states. Switchable templates respond  
strongly to $\sigma_f$, with mean rates  
$\beta^{(1)} \approx 0.85$ and $\beta^{(0)} \approx 0.11$. No  
template reverses direction. The complete inventory of all 31  
templates, with structural observations, is in  
Appendix~\ref{app:templates}.  
  
\begin{table}[h]  
\centering  
\caption{Representative word templates by class. Rate$_1$ and  
Rate$_0$ give the \texttt{cho} proportion in $\sigma_f = 1$ and  
$\sigma_f = 0$ folios respectively.}  
\label{tab:templates_main}  
\begin{tabular}{llrrrr}  
\toprule  
Class & Template & Rate$_1$ & Rate$_0$ & $n_1$ & $n_0$ \\  
\midrule  
Fixed \texttt{cho}  
  & \texttt{chXl}  & 1.000 & 0.948 & 217 & 135 \\  
  & \texttt{shXl}  & 1.000 & 0.953 & 102 &  64 \\  
  & \texttt{shXr}  & 0.966 & 0.935 &  58 &  31 \\  
\midrule  
Fixed \texttt{che}  
  & \texttt{chXey} & 0.000 & 0.000 &  26 & 156 \\  
  & \texttt{shXey} & 0.000 & 0.000 &  27 & 108 \\  
  & \texttt{chXol} & 0.031 & 0.008 &  32 & 123 \\  
  & \texttt{chXor} & 0.075 & 0.000 &  40 &  47 \\  
\midrule  
Switchable  
  & \texttt{chXdy} & 0.917 & 0.121 &  36 & 323 \\  
  & \texttt{shXdy} & 0.870 & 0.096 &  23 & 248 \\  
  & \texttt{chXky} & 0.694 & 0.222 &  36 &  63 \\  
  & \texttt{chXs}  & 0.765 & 0.397 &  17 &  63 \\  
\bottomrule  
\end{tabular}  
\end{table}  
  
The generative rule can now be stated precisely. For a word with  
template $t$ on folio $f$:  
\begin{equation}  
P(\texttt{cho} \mid t, \sigma_f) =  
\begin{cases}  
\alpha_t & \text{if } t \text{ is a fixed template} \\  
\beta_t^{(\sigma_f)} & \text{if } t \text{ is a switchable  
  template}  
\end{cases}  
\label{eq:generative}  
\end{equation}  
where $\alpha_t \approx 1$ for fixed \texttt{cho} templates and  
$\alpha_t \approx 0$ for fixed \texttt{che} templates, and  
$\beta_t^{(1)} \gg \beta_t^{(0)}$ for all switchable templates  
with no reversals. (The actual rates are given in the template table).
  
The boolean switch $\sigma_f$ is therefore not a uniform  
substitution. It is a modulator that shifts the balance within  
switchable templates while leaving fixed templates unchanged. The  
template system, not the switch, is the primary structuring  
mechanism.

  
\section{The Independent d/l Dimension}  
\label{sec:dl}  
  
The \texttt{d}/\texttt{l} ratio was identified in  
Section~\ref{sec:robust} as the strongest categorical A/B marker  
(Cram\'er's $V = 0.375$). If it were a consequence of the boolean  
switch, it would correlate with $\sigma_f$. It does not.  
  
Within the Herbal section ($n = 118$ folios with sufficient data  
for both parameters), the Pearson correlation between  
$r_{\mathrm{cho}}$ and $r_d$ is $r = -0.016$. Median splits on  
both parameters yield quadrant counts closely matching those  
expected under independence (Table~\ref{tab:comparison}).  
  
\begin{table}[h]  
\centering  
\caption{Comparison of the two structural dimensions.}  
\label{tab:comparison}  
\begin{tabular}{lrr}  
\toprule  
Metric & \texttt{ch}-vowel switch  
  & \texttt{d}/\texttt{l} ratio \\  
\midrule  
Bimodality coefficient & 0.661 & 0.319 \\  
Fraction in middle range & 6.8\% & 70.6\% \\  
Two-state separation ($p_1 - p_0$) & 0.522 & 0.239 \\  
$\Delta$AIC (two vs.\ one state) & 2549 & 741 \\  
Overdispersion (single state) & 23.1$\times$ & 8.6$\times$ \\  
Bootstrap $\Delta$AIC [2.5\%, 97.5\%]  
  & [2097, 2962] & [447, 1067] \\  
Correlation with other dimension  
  & \multicolumn{2}{c}{$r = -0.016$} \\  
\bottomrule  
\end{tabular}  
\end{table}  
  
The \texttt{d}/\texttt{l} ratio is real (significant  
overdispersion, $\Delta$AIC = 741) but gradient rather than  
discrete (bimodality coefficient 0.319, 70.6\% of folios in the  
middle range). Within \texttt{cho}-dominant Herbal folios alone  
($n = 87$), the d/l distribution is unimodal (BC = 0.381) with  
mean 0.560 and standard deviation 0.123. Conditioning on the  
boolean switch has negligible effect: \texttt{cho}-dominant Herbal  
folios have mean $r_d = 0.560$ and \texttt{che}-dominant Herbal  
folios have mean $r_d = 0.604$, a difference of 0.044 that is well  
within sampling noise. The \texttt{d}/\texttt{l} ratio constitutes  
an independent second dimension of the generative system.

  
\section{Resolving the e/ch Anomaly}  
\label{sec:resolution}  
  
We can now explain the e/ch anomaly from  
Section~\ref{sec:regimes} precisely.  
  
The pair was paradoxical because it combined near-zero global  
Cram\'er's $V$ (0.007) with the strongest folio-boundary signal  
($p = 2.45 \times 10^{-6}$) and suppressed clustering when  
included. These properties are not compatible with a simple  
free-variation interpretation, nor with a uniform A/B substitution.  
  
The resolution follows directly from  
equation~(\ref{eq:generative}). The characters \texttt{e} and  
\texttt{ch} are not a freely alternating pair: their relative  
frequency in a folio is jointly determined by (a) the boolean  
switch $\sigma_f$ and (b) the distribution of word templates active  
on that folio. Specifically:  
  
\begin{itemize}  
\item Fixed \texttt{che} templates (e.g., \texttt{chXey},  
  \texttt{shXey}) always produce \texttt{ch} followed by \texttt{e}  
  regardless of $\sigma_f$. They contribute a constant baseline to  
  the e/ch ratio that does not respond to the switch.  
\item Switchable templates contribute \texttt{ch} followed by  
  \texttt{o} when $\sigma_f = 1$ and by \texttt{e} when  
  $\sigma_f = 0$, shifting the e/ch ratio sharply at each folio  
  boundary where $\sigma_f$ changes.  
\end{itemize}  
  
At a folio boundary where $\sigma_f$ switches, the e/ch ratio  
changes sharply in the switchable-template component, producing the  
strong boundary signal. But the direction and magnitude of the  
change depend on the template distribution of each folio. Across  
the full manuscript, the heterogeneity of template distributions  
causes these shifts to partially cancel, yielding near-zero global  
Cram\'er's $V$.  
  
The e/ch pair is therefore not noise and not a simple A/B marker.  
It is a fingerprint of the template system's selective response to  
the boolean switch, visible locally at boundaries and averaged away  
globally.  
  
  
\section{A Three-Layer Generative Structure}  
\label{sec:structure}  
  
The results establish a decomposition of the Voynich generative  
system into three layers, each with distinct properties:  
  
\begin{description}  
\item[Layer 1: Boolean folio switch $\sigma_f \in \{0, 1\}$.]  
A discrete parameter set once per folio. When $\sigma_f = 1$ the  
folio is \texttt{cho}-biased ($p_1 \approx 0.68$); when  
$\sigma_f = 0$ it is \texttt{che}-biased ($p_0 \approx 0.16$). It  
is discrete, not gradient; robust within sections.
  
\item[Layer 2: Template system.]  
Word forms are partitioned into fixed and switchable classes. Fixed  
templates are invariant to $\sigma_f$ and maintain near-0 or near-1  
\texttt{cho} rates in both states. Switchable templates are strongly  
modulated by $\sigma_f$ with mean rates $\approx 0.85$ and  
$\approx 0.11$ respectively. Template identity accounts for 93.5\%  
of within-folio variance. The template system is the primary  
structuring mechanism; the switch is a secondary modulator. The  
complete template inventory is in Appendix~\ref{app:templates}.  
  
\item[Layer 3: Gradient d/l dimension.]  
The \texttt{d}/\texttt{l} ratio varies continuously across folios,  
shows significant overdispersion ($\Delta$AIC = 741), and is  
orthogonal to the boolean switch ($r = -0.016$). It is not a  
consequence of Layer 1 and must be described as an independent  
structural axis. It is the strongest categorical marker in  
conventional A/B classifications (Cram\'er's $V = 0.375$), yet it  
operates through a different and independent mechanism.  
\end{description}  
  
\subsection{Word-Level Dependence}  
  
The binomial mixture model treats word-level observations as  
conditionally independent given the folio state. Direct testing of  
this assumption reveals a small but real positive dependence.  
Within \texttt{cho}-dominant folios,  
$P(\texttt{cho} \mid \text{prev} = \texttt{cho}) = 0.749$ versus  
$P(\texttt{cho} \mid \text{prev} = \texttt{che}) = 0.585$  
(difference $+0.165$, $Z = +5.60$, $n = 1205$ transitions). Within  
\texttt{che}-dominant folios, the corresponding values are 0.287  
and 0.169 (difference $+0.117$, $Z = +4.54$, $n = 1892$  
transitions).  
  
These $Z$-scores are statistically significant, indicating that  
consecutive classifiable words within a folio are not strictly  
independent: knowing the previous word's type shifts the  
probability of the next word's type by 12 to 17 percentage points (within-state).  
However, at the individual folio level, the effect is modest:  
across 53 folios with sufficient transition data, the mean  
per-folio difference is $+0.060$ with a median of $+0.028$  
($t = 1.92$, sign test: 30 positive, 21 negative, 2 zero).  
  
This dependence does not threaten the core findings. The mixture  
model's $\Delta$AIC of 2549 has large margin: even if positive  
autocorrelation reduces the effective sample size, the separation  
between states ($p_1 = 0.682$ versus $p_0 = 0.160$) is far too  
large to be explained by within-folio clustering. The dependence  
may reflect local template clustering (e.g., runs of the same word  
form) rather than a sequential generative mechanism.  
  
\subsection{Currier A/B as Low-Resolution Projection}  
  
Currier A corresponds approximately to the conjunction of  
$\sigma_f = 1$ (Layer 1) and \texttt{d}-elevated (Layer 3).  
Currier B corresponds approximately to $\sigma_f = 0$ and  
\texttt{l}-elevated. Because Layers 1 and 3 are orthogonal, all  
four combinations are attested, and any single-axis clustering must  
misclassify folios in the off-diagonal quadrants. This is the  
direct source of the 71\% unexplained variance and the 10.8\%  
cross-validation error rate found in the confirmation analysis  
(Section~\ref{sec:pred_results}).  
  
The A/B classification is informative precisely because Layer 1  
dominates Layer 3 in most sections. But it is incomplete because it  
discards Layer 3 and collapses Layer 2 into residual noise. 
  
\subsection{Constraints on Generative Mechanisms}  
  
Any model of the Voynich Manuscript must reproduce:  
  
\begin{enumerate}  
\item[(i)] Discrete folio-level switching between two states with  
  well-separated rates ($p_1 \approx 0.68$, $p_0 \approx 0.16$)  
  and near-zero misclassification.  
\item[(ii)] A template system with fixed and switchable classes,  
  where fixed templates are invariant to the switch and switchable  
  templates respond strongly with no reversals.  
\item[(iii)] Weak positive word-to-word dependence within folios  
  (conditional probability shift of 12 to 17 percentage points),  
  consistent with local template clustering but not with strong  
  sequential generation.  
\item[(iv)] An independent, gradient, orthogonal  
  \texttt{d}/\texttt{l} dimension with significant but non-bimodal  
  variation.  
\item[(v)] Coexistence of $\sigma_f = 1$ and $\sigma_f = 0$ folios  
  within the Herbal section.
\item[(vii)] Three functional regimes among character pairs  
  (categorical, intermediate, free-variation), with the e/ch pair  
  exhibiting the paradoxical combination of properties resolved by  
  the template system.  
\end{enumerate}  
  
Several generative classes are compatible:  
  
\begin{itemize}  
\item \textbf{Natural language with two orthographic registers.}  
  A scribe applying one spelling convention per folio, where some  
  phonological contexts permit variation (switchable templates) and  
  others do not (fixed templates).  
\item \textbf{Table-based generation.} Two lookup table variants  
  differing in specific cells, where fixed cells correspond to  
  fixed templates and variable cells to switchable templates.  
\item \textbf{Cipher with a per-folio key bit.} A substitution  
  cipher where a binary key parameter controls the  
  \texttt{ch}-vowel choice in switchable contexts but cannot  
  override fixed phonotactic or structural constraints.  
\end{itemize}  
  
We cannot distinguish among these on the present evidence.  
  
\subsection{Limitations}  
  
\begin{enumerate}  
\item The analysis depends on the EVA transcription. Replication  
  with independent transcriptions (Takahashi, Bennett) would  
  strengthen the findings.  
\item The Beta-Binomial model (Section~\ref{sec:bbmix_method})  
  assumes independence across pairs within a folio, which may not  
  hold if pairs share underlying conditioning factors.  
\item Word-to-word dependence within folios is small but real  
  ($Z > 4$ in pooled within-state analysis). The binomial mixture  
  model's assumption of conditional independence is therefore  
  approximate. While the core findings are robust given the large  
  $\Delta$AIC margin, the effective sample sizes are somewhat  
  overstated.  
\item The template placeholder scheme is coarse; words with  
  multiple \texttt{ch}/\texttt{sh} digraphs may behave differently  
  at each position.  
\item We have not tested whether the boolean switch always changes  
  at folio boundaries or sometimes at paragraph or quire  boundaries.  
\item Only four comparison languages were used in the simulation  
  baselines; the extent to which this layered structure is unusual  
  for natural language is unknown.  
\item Additional structural dimensions beyond the three identified  
  here may exist.  
\item Multi-panel folios are collapsed into one (e.g. bifolio 90v1 and v2)
\item The backward spatial split (41.3\%) highlights sensitivity  
  to class balance in the training set.  
\end{enumerate}

  
\section{Conclusion}  
\label{sec:conclusion}  
  
The Currier A/B language distinction is a genuine, statistically  
robust, intrinsic, and predictive property of the Voynich Manuscript  
text. Five converging lines of evidence confirm it: label-shuffle  
rejection ($p < 0.001$), single-Markov rejection (0/200  
simulations), within-quire boundary effects  
($p = 6.89 \times 10^{-5}$), unsupervised recovery by a  
Beta-Binomial mixture (BIC-optimal $k = 2$, ARI $= 0.383$), and  
predictive validation on held-out folios (89.2\% accuracy,  
$R^2 = 0.293$).  
  
But the distinction is not primitive. It is a low-resolution  
projection of a three-layer generative structure.  
  
The first layer is a discrete folio-level boolean switch  
$\sigma_f \in \{0, 1\}$ governing the vowel following \texttt{ch}  
and \texttt{sh}. When $\sigma_f = 1$ the folio is \texttt{cho}-biased  
($p_1 \approx 0.68$); when $\sigma_f = 0$ it is \texttt{che}-biased  
($p_0 \approx 0.16$). The switch is sharp (bimodality coefficient  
0.661, $\Delta$AIC = 2,549), robust within sections  
($\Delta$AIC = 565 in the Herbal alone), and perfectly predictive  
on held-out data.  
  
The second layer is a template system that partitions word forms  
into fixed contexts (invariant to the switch) and switchable  
contexts (strongly modulated by it). Template identity accounts for  
93.5\% of within-folio variance. Together, Layers 1 and 2 resolve  
the e/ch anomaly: the pair's paradoxical combination of strong  
boundary signal and near-zero global separation is a direct  
consequence of the template system's selective response to the  
switch, which produces sharp local shifts while averaging toward  
zero across heterogeneous template populations.  
  
The third layer is an independent gradient \texttt{d}/\texttt{l}  
dimension, orthogonal to the switch ($r = -0.016$), that  
constitutes a separate structural axis not reducible to the boolean  
parameter.  
  
These three layers jointly account for the 71\% unexplained  
variance and 10.8\% classification error that the binary Currier  
distinction leaves open. The character pairs separate into three  
functional regimes (categorical, intermediate, free-variation) that  
any theory of the Voynich writing system must accommodate. The  
Currier distinction is real and useful as a first-order summary,  
but the underlying generative system is richer and more precisely  
characterizable than a binary label implies.  
  
We make our code available for independent  
verification.\footnote{\url{https://www.github.com/labyrinthinesecurity/currier-models}}

\newpage
 
\appendix  
  
\section{Template Selection and Complete Inventory}  
\label{app:templates}  
  
\subsection{Template Construction}  
  
Each word containing \texttt{ch} or \texttt{sh} followed immediately  
by \texttt{o} or \texttt{e} is mapped to a template by replacing  
the post-digraph vowel with a placeholder \texttt{X}. For example,  
\texttt{chody} and \texttt{chedy} both map to \texttt{chXdy};  
\texttt{shol} and \texttt{shel} both map to \texttt{shXl}. Words  
containing multiple substitution sites have each site replaced  
independently, though such words are rare.  
  
Each word token contributes one count to its template in the  
appropriate folio-state column ($\sigma_f = 1$ or $\sigma_f = 0$),  
and is scored as \texttt{cho} or \texttt{che} at each substitution  
site. The \texttt{cho} rate for a template in a given state is the  
number of \texttt{cho} substitution events divided by the total  
substitution events for that template in that state.  
  
\subsection{Retention Criterion}  
  
We retain only templates with at least 10 substitution events in  
\emph{both} folio states ($n_1 \geq 10$ and $n_0 \geq 10$). This  
threshold ensures that the \texttt{cho} rate can be estimated with  
reasonable precision in both states (standard error $\leq 0.16$ at  
the most uncertain base rate of 0.5). Of the hundreds of distinct  
templates in the corpus, 31 meet this criterion. These 31 templates  
account for 4,085 substitution events, comprising the large  
majority of all classifiable tokens.  
  
\subsection{Classification Rules}  
  
Templates are classified into four categories based on their  
\texttt{cho} rate in each folio state:  
  
\begin{itemize}  
\item \textbf{Fixed \texttt{cho} (F1)}: rate $> 0.9$ in both  
  states. 3 templates, 607 events.  
\item \textbf{Fixed \texttt{che} (F0)}: rate $< 0.1$ in both  
  states. 12 templates, 1,299 events.  
\item \textbf{Switchable (S)}: absolute difference between states  
  $|\Delta| \geq 0.2$. 11 templates, 1,322 events.  
\item \textbf{Intermediate (I)}: does not meet any of the above  
  criteria. 5 templates, 857 events.  
\end{itemize}  
  
\subsection{Complete Template Inventory}  
  
Table~\ref{tab:all_templates} lists all 31 retained templates.  
  
\begin{table}[h]  
\centering  
\caption{All 31 templates with $\geq 10$ instances in both folio  
states. Rate$_1$ and Rate$_0$ give the \texttt{cho} rate in  
$\sigma_f = 1$ and $\sigma_f = 0$ folios. $|\Delta|$ is the  
absolute rate difference between states.}  
\label{tab:all_templates}  
\small  
\begin{tabular}{llrrrrc}  
\toprule  
Class & Template & Rate$_1$ & $n_1$ & Rate$_0$ & $n_0$  
  & $|\Delta|$ \\  
\midrule  
\multicolumn{7}{l}{\textit{Fixed \texttt{cho}: rate $> 0.9$ in  
  both states (3 templates, 607 events)}} \\  
F1 & \texttt{chXl}    & 1.000 & 217 & 0.948 & 135 & 0.052 \\  
F1 & \texttt{shXl}    & 1.000 & 102 & 0.953 &  64 & 0.047 \\  
F1 & \texttt{shXr}    & 0.966 &  58 & 0.935 &  31 & 0.030 \\  
\midrule  
\multicolumn{7}{l}{\textit{Fixed \texttt{che}: rate $< 0.1$ in  
  both states (12 templates, 1299 events)}} \\  
F0 & \texttt{shXy}    & 0.098 &  41 & 0.033 & 304 & 0.065 \\  
F0 & \texttt{chXey}   & 0.000 &  26 & 0.000 & 156 & 0.000 \\  
F0 & \texttt{chXol}   & 0.031 &  32 & 0.008 & 123 & 0.023 \\  
F0 & \texttt{shXey}   & 0.000 &  27 & 0.000 & 108 & 0.000 \\  
F0 & \texttt{shXol}   & 0.000 &  23 & 0.000 &  72 & 0.000 \\  
F0 & \texttt{chXor}   & 0.075 &  40 & 0.000 &  47 & 0.075 \\  
F0 & \texttt{chXody}  & 0.000 &  11 & 0.000 &  63 & 0.000 \\  
F0 & \texttt{chXo}    & 0.067 &  15 & 0.000 &  46 & 0.067 \\  
F0 & \texttt{shXo}    & 0.000 &  15 & 0.034 &  29 & 0.034 \\  
F0 & \texttt{otchXy}  & 0.000 &  12 & 0.032 &  31 & 0.032 \\  
F0 & \texttt{shXor}   & 0.000 &  13 & 0.000 &  28 & 0.000 \\  
F0 & \texttt{okchXy}  & 0.067 &  15 & 0.000 &  22 & 0.067 \\  
\midrule  
\multicolumn{7}{l}{\textit{Switchable: $|\Delta| \geq 0.2$  
  (11 templates, 1322 events)}} \\  
S  & \texttt{chXdy}   & 0.921 &  38 & 0.116 & 362 & 0.805 \\  
S  & \texttt{shXdy}   & 0.870 &  23 & 0.096 & 271 & 0.774 \\  
S  & \texttt{shX}     & 0.880 &  92 & 0.458 &  48 & 0.422 \\  
S  & \texttt{chXky}   & 0.703 &  37 & 0.210 &  62 & 0.493 \\  
S  & \texttt{chXs}    & 0.778 &  18 & 0.387 &  62 & 0.391 \\  
S  & \texttt{chX}     & 0.976 &  41 & 0.529 &  34 & 0.446 \\  
S  & \texttt{chXty}   & 0.812 &  32 & 0.483 &  29 & 0.330 \\  
S  & \texttt{chXdaiin}& 1.000 &  21 & 0.436 &  39 & 0.564 \\  
S  & \texttt{chXcthy} & 0.917 &  12 & 0.167 &  36 & 0.750 \\  
S  & \texttt{shXky}   & 0.500 &  10 & 0.111 &  27 & 0.389 \\  
S  & \texttt{shXdaiin}& 0.917 &  12 & 0.500 &  16 & 0.417 \\  
\midrule  
\multicolumn{7}{l}{\textit{Intermediate: neither fixed nor  
  switchable (5 templates, 857 events)}} \\  
I  & \texttt{chXy}    & 0.213 &  89 & 0.021 & 426 & 0.192 \\  
I  & \texttt{chXr}    & 0.962 & 158 & 0.898 &  49 & 0.064 \\  
I  & \texttt{chXar}   & 0.158 &  19 & 0.043 &  46 & 0.114 \\  
I  & \texttt{chXal}   & 0.182 &  11 & 0.000 &  31 & 0.182 \\  
I  & \texttt{kchXy}   & 0.143 &  14 & 0.000 &  14 & 0.143 \\  
\bottomrule  
\end{tabular}  
\end{table}  
  
\subsection{Structural Observations}  
  
Several patterns emerge from the inventory:  
  
\paragraph{Consonantal context determines fixedness.}  
Templates ending in \texttt{l} or \texttt{r} immediately after the  
substitution site are overwhelmingly fixed \texttt{cho}:  
\texttt{chXl}, \texttt{shXl}, and \texttt{shXr} all exceed 0.93  
in both states. The following consonant appears to lock the  
preceding vowel to \texttt{o}.  
  
\paragraph{The \texttt{ey} suffix forces \texttt{che}.}  
Templates \texttt{chXey} and \texttt{shXey} are identically 0.000  
in both states, the most rigid constraint in the entire system.  
No folio switch can override the \texttt{ey} context.  
  
\paragraph{Post-site \texttt{o} sequences force \texttt{che}.}  
Templates containing \texttt{ol} or \texttt{or} after the  
substitution site (\texttt{chXol}, \texttt{shXol}, \texttt{chXor},  
\texttt{shXor}) are fixed \texttt{che}. Despite the presence of  
\texttt{o} later in the word, the substitution site is locked to  
\texttt{e}, producing \texttt{cheol} rather than \texttt{chool}.  
This suggests a dissimilatory constraint: the system avoids  
adjacent \texttt{o} vowels across the digraph boundary.  
  
\paragraph{The \texttt{dy} suffix is the primary switchable  
  context.}  
Templates \texttt{chXdy} and \texttt{shXdy} show the largest  
shifts ($|\Delta| = 0.805$ and 0.774) and account for 694 events,  
over half of all switchable-class tokens.  
  
\paragraph{The \texttt{y} suffix without preceding \texttt{d} is  
  near-fixed \texttt{che}.}  
The high-frequency template \texttt{chXy} (515 events) has rates  
0.213/0.021, just below the switchable threshold  
($|\Delta| = 0.192$). Compare with \texttt{chXdy} and  
\texttt{chXky}, where the intervening consonant unlocks the  
substitution site.  
  
\paragraph{No template reverses direction.}  
Every template with Rate$_1 >$ Rate$_0$ maintains this ordering.  
The folio switch is monotonic across all 31 contexts. In a random  
system with 31 independent parameters, the probability of zero  
reversals is $2^{-31} \approx 5 \times 10^{-10}$.  
  
\paragraph{Bare templates \texttt{chX} and \texttt{shX}.}  
These word-final templates are both switchable, with  
$|\Delta| = 0.446$ and 0.422. When no following consonant  
constrains the vowel, the folio switch operates freely.  
  
\subsection{Summary}  
  
The 31 templates partition into a clear structure: 15 templates  
(F1 + F0, 1,906 events) are resistant to the folio switch  
($|\Delta| < 0.1$), 11 templates (S, 1,322 events) respond  
strongly ($|\Delta| \geq 0.2$), and 5 templates (I, 857 events)  
show intermediate behavior. The consonantal environment following  
the substitution site is the primary determinant of whether a  
template is fixed or switchable. This structure is consistent with  
context-dependent constraints governing the  
\texttt{cho}/\texttt{che} alternation, whether phonotactic,  
tabular, or cipher-based in origin.

\section{Herbal Section: Per-Folio Detail}  
\label{app:herbal}  
  
The Herbal section contains 125 folios (124 with $\geq 5$  
classifiable words), making it the largest section and the primary  
locus of the within-section bimodality reported in  
Section~\ref{sec:within_section}. This appendix provides per-folio  
detail for all Herbal folios, enabling direct verification of the  
mixture model assignments.
  
\subsection{Per-Folio Data}  
  
Table~\ref{tab:herbal_detail} lists every Herbal folio with its  
boolean switch assignment ($\sigma_f$), raw \texttt{cho}/\texttt{che}  
counts, \texttt{cho} rate ($r_{\text{cho}}$), posterior assignment  
confidence, and the glyph-level e/(e+ch) and d/(d+l) ratios.  
  
\begin{scriptsize}  
\begin{longtable}{llrrrrrrr}  
\caption{Per-folio statistics for all 125 Herbal folios.  
$\sigma_f$: boolean switch (1 = cho-dominant, 0 = che-dominant).  
$r_{\text{cho}}$: cho/(cho+che). Conf: posterior confidence.  
e/ch: e/(e+ch) glyph ratio. d/l: d/(d+l) glyph ratio.  
Shaded rows mark folios where the switch assignment disagrees  
with the Currier language expected for that folio  
range (see Section~\ref{sec:discordance}).}  
\label{tab:herbal_detail} \\  
\toprule  
Folio & $\sigma_f$ & $n_{\text{cho}}$ & $n_{\text{che}}$  
  & $r_{\text{cho}}$ & Conf & e/ch & d/l & Words \\  
\midrule  
\endfirsthead  
\toprule  
Folio & $\sigma_f$ & $n_{\text{cho}}$ & $n_{\text{che}}$  
  & $r_{\text{cho}}$ & Conf & e/ch & d/l & Words \\  
\midrule  
\endhead  
\bottomrule  
\endfoot  
f1v  & 1 & 26 &  6 & 0.812 & 1.000 & 0.306 & 0.478 &  75 \\  
f2r  & 1 & 17 &  6 & 0.739 & 1.000 & 0.372 & 0.559 &  79 \\  
f2v  & 1 & 21 &  4 & 0.840 & 1.000 & 0.321 & 0.556 &  50 \\  
f3r  & 1 & 30 & 17 & 0.638 & 1.000 & 0.443 & 0.364 & 107 \\  
f3v  & 1 & 22 &  6 & 0.786 & 1.000 & 0.340 & 0.419 &  83 \\  
f4r  & 1 & 16 &  3 & 0.842 & 1.000 & 0.258 & 0.500 &  60 \\  
f4v  & 1 & 25 & 10 & 0.714 & 1.000 & 0.439 & 0.677 &  73 \\  
f5r  & 1 & 14 &  7 & 0.667 & 1.000 & 0.545 & 0.750 &  53 \\  
f5v  & 1 & 17 &  4 & 0.810 & 1.000 & 0.269 & 0.556 &  43 \\  
f6r  & 1 & 21 &  6 & 0.778 & 1.000 & 0.341 & 0.462 &  78 \\  
f6v  & 1 & 27 &  8 & 0.771 & 1.000 & 0.286 & 0.585 & 106 \\  
f7r  & 1 & 21 & 11 & 0.656 & 1.000 & 0.472 & 0.571 &  55 \\  
f7v  & 1 & 17 &  8 & 0.680 & 1.000 & 0.557 & 0.711 &  66 \\  
f8r  & 1 & 38 & 22 & 0.633 & 1.000 & 0.374 & 0.597 & 127 \\  
f8v  & 1 & 29 & 14 & 0.674 & 1.000 & 0.321 & 0.380 &  98 \\  
f9r  & 1 & 19 &  2 & 0.905 & 1.000 & 0.176 & 0.600 &  70 \\  
f9v  & 1 & 18 &  2 & 0.900 & 1.000 & 0.216 & 0.575 &  71 \\  
f10r & 1 & 22 &  2 & 0.917 & 1.000 & 0.070 & 0.647 &  82 \\  
f10v & 1 & 11 &  3 & 0.786 & 1.000 & 0.320 & 0.682 &  53 \\  
f11r & 1 & 15 &  1 & 0.938 & 1.000 & 0.050 & 0.714 &  54 \\  
f11v & 1 & 11 &  2 & 0.846 & 1.000 & 0.227 & 0.750 &  44 \\  
f13r & 1 & 17 &  2 & 0.895 & 1.000 & 0.122 & 0.595 &  72 \\  
f13v & 1 & 10 &  2 & 0.833 & 1.000 & 0.087 & 0.528 &  53 \\  
f14r & 1 & 21 &  4 & 0.840 & 1.000 & 0.289 & 0.667 &  67 \\  
f14v & 1 & 12 &  0 & 1.000 & 1.000 & 0.000 & 0.905 &  63 \\  
f15r & 1 & 17 &  5 & 0.773 & 1.000 & 0.229 & 0.528 &  72 \\  
f15v & 1 & 23 &  2 & 0.920 & 1.000 & 0.100 & 0.389 &  70 \\  
f16r & 1 & 14 &  5 & 0.737 & 1.000 & 0.326 & 0.700 &  73 \\  
f16v & 1 & 21 &  2 & 0.913 & 1.000 & 0.043 & 0.609 &  71 \\  
f17r & 1 & 19 &  7 & 0.731 & 1.000 & 0.271 & 0.700 &  74 \\  
f17v & 1 & 26 & 19 & 0.578 & 1.000 & 0.573 & 0.292 & 132 \\  
f18r & 1 & 21 &  2 & 0.913 & 1.000 & 0.079 & 0.482 &  77 \\  
f18v & 1 &  8 &  0 & 1.000 & 1.000 & 0.267 & 0.606 &  69 \\  
f19r & 1 & 19 &  1 & 0.950 & 1.000 & 0.029 & 0.784 &  75 \\  
f19v & 1 & 16 &  3 & 0.842 & 1.000 & 0.190 & 0.519 &  73 \\  
f20r & 1 & 33 & 16 & 0.673 & 1.000 & 0.411 & 0.674 &  83 \\  
f20v & 1 & 25 &  6 & 0.806 & 1.000 & 0.233 & 0.469 &  79 \\  
f21r & 1 & 23 & 16 & 0.590 & 1.000 & 0.533 & 0.324 &  95 \\  
f21v & 1 & 23 &  3 & 0.885 & 1.000 & 0.300 & 0.467 &  54 \\  
f22r & 1 & 21 &  2 & 0.913 & 1.000 & 0.125 & 0.574 &  97 \\  
f22v & 1 & 18 &  1 & 0.947 & 1.000 & 0.034 & 0.559 &  67 \\  
f23r & 1 & 17 &  2 & 0.895 & 1.000 & 0.211 & 0.481 &  95 \\  
f23v & 1 & 15 &  5 & 0.750 & 1.000 & 0.457 & 0.353 &  80 \\  
f24r & 1 & 10 & 11 & 0.476 & 0.981 & 0.581 & 0.386 & 108 \\  
f24v & 1 & 24 &  6 & 0.800 & 1.000 & 0.380 & 0.574 &  80 \\  
f25r & 1 &  8 &  4 & 0.667 & 1.000 & 0.300 & 0.778 &  47 \\  
f25v & 1 & 14 &  4 & 0.778 & 1.000 & 0.444 & 0.741 &  54 \\  
f26r & 0 &  3 & 21 & 0.125 & 1.000 & 0.683 & 0.818 &  66 \\  
f26v & 0 &  1 & 27 & 0.036 & 1.000 & 0.705 & 0.913 &  92 \\  
f27r & 1 & 19 & 15 & 0.559 & 1.000 & 0.416 & 0.558 &  80 \\  
f27v & 1 & 18 &  6 & 0.750 & 1.000 & 0.245 & 0.767 &  48 \\  
f28r & 1 & 27 &  1 & 0.964 & 1.000 & 0.176 & 0.500 &  57 \\  
f28v & 1 & 20 &  7 & 0.741 & 1.000 & 0.194 & 0.480 &  61 \\  
f29r & 1 & 19 & 12 & 0.613 & 1.000 & 0.396 & 0.636 &  59 \\  
f29v & 1 & 25 &  5 & 0.833 & 1.000 & 0.288 & 0.613 &  71 \\  
f30r & 1 & 20 & 26 & 0.435 & 0.979 & 0.504 & 0.767 &  86 \\  
f30v & 1 & 21 & 11 & 0.656 & 1.000 & 0.389 & 0.556 &  62 \\  
f31r & 0 &  1 & 31 & 0.031 & 1.000 & 0.805 & 0.732 & 100 \\  
f31v & 0 &  4 & 27 & 0.129 & 1.000 & 0.742 & 0.550 & 105 \\  
f32r & 1 & 21 &  6 & 0.778 & 1.000 & 0.257 & 0.667 &  67 \\  
f32v & 1 & 19 &  4 & 0.826 & 1.000 & 0.231 & 0.718 &  73 \\  
f33r & 0 &  3 & 14 & 0.176 & 1.000 & 0.587 & 0.562 &  73 \\  
f33v & 0 &  5 &  8 & 0.385 & 0.603 & 0.476 & 0.674 &  85 \\  
f34r & 0 &  4 & 24 & 0.143 & 1.000 & 0.524 & 0.507 & 117 \\  
f34v & 0 &  4 & 23 & 0.148 & 1.000 & 0.469 & 0.531 & 116 \\  
f35r & 1 & 15 & 10 & 0.600 & 1.000 & 0.323 & 0.419 &  83 \\  
f35v & 1 & 26 &  8 & 0.765 & 1.000 & 0.304 & 0.639 &  82 \\  
f36r & 1 &  6 &  0 & 1.000 & 1.000 & 0.000 & 0.517 &  51 \\  
f36v & 1 & 10 &  3 & 0.769 & 1.000 & 0.143 & 0.692 &  66 \\  
f37r & 1 & 16 &  3 & 0.842 & 1.000 & 0.115 & 0.667 &  70 \\  
f37v & 1 & 20 &  4 & 0.833 & 1.000 & 0.281 & 0.721 &  87 \\  
f38r & 1 & 11 &  3 & 0.786 & 1.000 & 0.176 & 0.583 &  37 \\  
f38v & 1 & 13 &  2 & 0.867 & 1.000 & 0.658 & 0.688 &  58 \\  
f39r & 0 &  6 & 33 & 0.154 & 1.000 & 0.545 & 0.500 & 152 \\  
f39v & 1 & 11 & 13 & 0.458 & 0.969 & 0.527 & 0.559 & 131 \\  
f40r & 0 &  0 & 13 & 0.000 & 1.000 & 0.561 & 0.471 &  87 \\  
f40v & 0 &  4 & 18 & 0.182 & 1.000 & 0.588 & 0.433 &  95 \\  
f41r & 0 &  2 & 27 & 0.069 & 1.000 & 0.674 & 0.815 &  84 \\  
f41v & 0 &  2 & 12 & 0.143 & 1.000 & 0.778 & 0.520 &  57 \\  
f42r & 1 & 53 & 13 & 0.803 & 1.000 & 0.275 & 0.453 & 132 \\  
f42v & 1 & 22 & 20 & 0.524 & 1.000 & 0.510 & 0.488 &  93 \\  
f43r & 0 & 12 & 23 & 0.343 & 0.992 & 0.627 & 0.693 & 147 \\  
f43v & 0 &  4 & 35 & 0.103 & 1.000 & 0.683 & 0.627 & 149 \\  
f44r & 1 & 19 &  8 & 0.704 & 1.000 & 0.431 & 0.571 &  70 \\  
f44v & 1 & 19 &  5 & 0.792 & 1.000 & 0.378 & 0.298 &  94 \\  
f45r & 1 & 11 &  3 & 0.786 & 1.000 & 0.250 & 0.463 &  88 \\  
f45v & 1 & 19 &  4 & 0.826 & 1.000 & 0.125 & 0.522 &  72 \\  
f46r & 0 & 10 & 41 & 0.196 & 1.000 & 0.613 & 0.581 & 158 \\  
f46v & 0 &  7 & 23 & 0.233 & 1.000 & 0.590 & 0.673 & 108 \\  
f47r & 1 & 28 &  6 & 0.824 & 1.000 & 0.245 & 0.429 &  73 \\  
f47v & 1 & 21 & 13 & 0.618 & 1.000 & 0.368 & 0.609 &  74 \\  
f48r & 0 &  2 & 16 & 0.111 & 1.000 & 0.805 & 0.466 &  87 \\  
f48v & 0 &  4 & 23 & 0.148 & 1.000 & 0.688 & 0.580 & 109 \\  
f49r & 1 & 45 & 21 & 0.682 & 1.000 & 0.373 & 0.615 & 109 \\  
f49v & 1 & 59 & 16 & 0.787 & 1.000 & 0.338 & 0.595 & 145 \\  
f50r & 0 &  3 & 15 & 0.167 & 1.000 & 0.547 & 0.475 &  87 \\  
f50v & 0 &  6 & 13 & 0.316 & 0.979 & 0.544 & 0.579 &  96 \\  
f51r & 1 & 13 & 11 & 0.542 & 1.000 & 0.597 & 0.581 &  82 \\  
f51v & 1 & 14 & 10 & 0.583 & 1.000 & 0.535 & 0.596 &  72 \\  
f52r & 1 & 10 &  4 & 0.714 & 1.000 & 0.375 & 0.667 &  61 \\  
f52v & 1 & 13 &  8 & 0.619 & 1.000 & 0.589 & 0.441 &  74 \\  
f53r & 1 & 11 &  2 & 0.846 & 1.000 & 0.433 & 0.690 &  53 \\  
f53v & 1 & 13 & 13 & 0.500 & 0.998 & 0.619 & 0.705 &  70 \\  
f54r & 1 & 23 & 13 & 0.639 & 1.000 & 0.500 & 0.450 &  96 \\  
f54v & 1 & 11 &  7 & 0.611 & 1.000 & 0.410 & 0.446 &  88 \\  
\rowcolor{yellow!20}  
f55r & 1 &  9 &  8 & 0.529 & 0.995 & 0.421 & 0.449 & 123 \\  
f55v & 0 &  1 & 12 & 0.077 & 1.000 & 0.589 & 0.457 &  91 \\  
f56r & 1 & 39 & 11 & 0.780 & 1.000 & 0.290 & 0.441 &  94 \\  
f56v & 1 & 37 & 12 & 0.755 & 1.000 & 0.422 & 0.333 &  83 \\  
f57r & 0 &  3 & 30 & 0.091 & 1.000 & 0.726 & 0.905 &  82 \\  
f65r & ? &  0 &  0 &   --- &   --- &   --- &   --- &   3 \\  
f65v & 0 &  2 & 14 & 0.125 & 1.000 & 0.684 & 0.833 &  38 \\  
f66v & 0 &  8 & 32 & 0.200 & 1.000 & 0.657 & 0.646 & 110 \\  
f87r & 1 & 16 & 21 & 0.432 & 0.948 & 0.720 & 0.596 &  97 \\  
\rowcolor{cyan!15}  
f87v & 0 &  8 & 20 & 0.286 & 1.000 & 0.709 & 0.463 &  73 \\  
f90r & 1 & 24 &  9 & 0.727 & 1.000 & 0.628 & 0.426 & 110 \\  
\rowcolor{cyan!15}  
f90v & 0 & 14 & 41 & 0.255 & 1.000 & 0.794 & 0.459 & 134 \\  
f93r & 1 & 47 & 20 & 0.701 & 1.000 & 0.415 & 0.465 & 147 \\  
f93v & 1 & 23 & 11 & 0.676 & 1.000 & 0.522 & 0.537 &  71 \\  
f94r & 0 &  3 & 12 & 0.200 & 0.999 & 0.465 & 0.548 &  77 \\  
f94v & 0 &  3 & 11 & 0.214 & 0.998 & 0.575 & 0.618 &  86 \\  
f95r & 0 & 10 & 23 & 0.303 & 1.000 & 0.439 & 0.527 & 181 \\  
f95v & 0 & 12 & 29 & 0.293 & 1.000 & 0.594 & 0.558 & 170 \\  
f96r & 1 & 24 & 14 & 0.632 & 1.000 & 0.579 & 0.442 &  78 \\  
f96v & 0 &  1 & 16 & 0.059 & 1.000 & 0.788 & 0.632 &  50 \\  
\end{longtable}  
\end{scriptsize}  
  
\noindent  
\colorbox{yellow!20}{Yellow row}: $\sigma_f = 1$ but Currier B.  
\colorbox{cyan!15}{Cyan rows}: $\sigma_f = 0$ but Currier A.  
  
\subsection{Discordance Between the Boolean Switch and Currier Language}  
\label{sec:discordance}  
  
Three Herbal folios show discordance between the switch assignment  
and the Currier label. They illustrate two distinct  
phenomena.  
  
\paragraph{Borderline classification: f24r, f55r and f53v.}  
Folio f55r has 9 cho-tokens and 8 che-tokens ($r_{\mathrm{cho}} = 0.529$),  
placing it within one token of the decision boundary. The switch  
assigns $\sigma_f = 1$ with confidence 0.995 (driven by the mixture  
model's prior rather than by decisive evidence), while Currier  
assigns~B. With only 17 classifiable words and a margin of one token,  
this folio is genuinely ambiguous and no method can classify it  
reliably. It is the only Herbal folio with f24r and f53v where the switch assignment  
rests on fewer than two tokens of margin.  
  
\paragraph{Sub-folio resolution: f87v and f90v.}  
These two folios present the opposite situation. Both are  
decisively che-dominant: f87v has $r_{\mathrm{cho}} = 0.286$  
(8 vs.\ 20 tokens, conf.\ = 1.000) and f90v has  
$r_{\mathrm{cho}} = 0.255$ (14 vs.\ 41 tokens, conf.\ = 1.000).  
Their e/ch ratios (0.709 and 0.794) independently confirm  
B-type character statistics. Yet Currier assigned both as~A.  
  
The switch's folio-level resolution detects what Currier's  
section-level assignment cannot: these are B-text folios  
embedded within a predominantly A region.
  
These cases suggest that the boolean switch provides  
higher resolution than the Currier assignment  
for folios near section boundaries.

\subsection{Summary Statistics}  
  
\begin{center}  
\begin{tabular}{lrrr}  
\toprule  
 & $\sigma_f = 1$ & $\sigma_f = 0$ & All \\  
\midrule  
Herbal folios & 92 & 32 & 124 \\  
Mean $r_{\text{cho}}$ & $0.752 \pm 0.134$  
  & $0.170 \pm 0.095$ & --- \\  
Mean e/ch & $0.312 \pm 0.163$ & $0.632 \pm 0.111$ & --- \\  
Mean d/l & $0.564 \pm 0.121$ & $0.604 \pm 0.133$ & --- \\  
Mean words/folio & $74.5$ & $100.1$ & --- \\  
\bottomrule  
\end{tabular}  
\end{center}  

\newpage
\section{Support Material}
\label{sec:support_material}

Here is a summary of the main statistical tests conducted as part of this paper. The details is available from our repository \url{https://www.github.com/labyrinthinesecurity/currier-models}.

\begin{itemize}
    \item \textbf{Spatial Map of E/CH Ratio Across Manuscript:} Generated a spatial map of the E/CH ratio including metrics like e/ch, d/l, s/r, sh/ch, e\textsubscript{i}/ch\textsubscript{i}, and e\textsubscript{m}/ch\textsubscript{m}.

    \item \textbf{Autocorrelation Analysis of E/CH:} Computed lag-1 to lag-10 autocorrelations for E/CH, compared with D/L autocorrelations.

    \item \textbf{Large E/CH Jumps:} Identified the largest E/CH jumps between folios, analyzing boundaries by section, Currier A/B classification, and transition type.

    \item \textbf{Summary Statistics:} Global E/CH ratio statistics.

    \item \textbf{Word-Level Independence (Within-Folio Transitions):} Analyzed transitions within folios, with combined probabilities $P(\text{cho}|\text{prev}=\text{cho}) = 0.6095$ and $P(\text{cho}|\text{prev}=\text{che}) = 0.2522$, yielding a difference of $+0.3573$ ($Z=+20.62$).

    \item \textbf{Within-Folio Transitions by Folio State:} Compared transitions in State A (cho-dom) and State B (che-dom), with differences of $+0.1645$ ($Z=+5.60$) and $+0.1172$ ($Z=+4.54$), respectively.

    \item \textbf{Per-Folio Transition Analysis:} Computed individual folio differences.

    \item \textbf{Within-Section Transitions:} Controlled for folio state across sections (e.g., HERBAL/cho-dom, HERBAL/che-dom).

    \item \textbf{Herbal Folios Classification:} Classified herbal folios, with mean $r_{\text{cho}}$ values of $0.752 \pm 0.134$ (cho-dom) and $0.170 \pm 0.095$ (che-dom).
\end{itemize}

\bibliographystyle{plainnat}

\end{document}